# Intrinsic structural and electronic properties of the Buffer Layer on Silicon Carbide unraveled by Density Functional Theory

Tommaso Cavallucci and Valentina Tozzini*

*NEST- Scuola Normale Superiore and Istituto Nanoscienze, Cnr, Piazza San Silvestro 12, 56127 Pisa, Italy*

**Abstract**

The buffer carbon layer obtained in the first instance by evaporation of Si from the Si-rich surfaces of silicon carbide (SiC) is often studied only as the intermediate to the synthesis of SiC supported graphene. In this work, we explore its intrinsic potentialities, addressing its structural and electronic properties by means Density Functional Theory. While the system of corrugation crests organized in a honeycomb super-lattice of nano-metric side returned by calculations is compatible with atomic microscopy observations, our work reveals some possible alternative symmetries, which might coexist in the same sample. The electronic structure analysis reveals the presence of an electronic gap of ~0.7eV. In-gap states are present, localized over the crests, while near-gap states reveal very different structure and space localization, being either bonding states or outward pointing p orbitals and unsaturated Si dangling bonds. On one hand, he presence of these interface states was correlated with the n-doping of the monolayer graphene subsequently grown on the buffer. On the other hand, the correlation between their chemical character and their space localization is likely to produce a differential reactivity towards specific functional groups with a spatial regular modulation at the nano-scale, opening perspectives for a finely controlled chemical functionalization.

## 1. Introduction

The thermal decomposition of silicon carbide (SiC) is a widely used technique to produce supported graphene[1]. Upon selective evaporation of Si from a Si-rich face of SiC (typically the 111 of 4C or the 0001 of hexagonal polytypes), excess carbon reconstruction produces in the first instance a honeycomb carbon layer covalently bound to the substrate called the buffer layer[2] (BL). Due to the non negligible residual amount of $sp^3$ hybridized sites, the BL is not graphene, but the latter can be obtained from it either by further evaporation of Si, producing the so called "epitaxial" graphene monolayer (EM), or by intercalation of hydrogen[3] or metals[4] underneath the buffer layer (Quasi Free Standing Monolayer Graphene, QFSMG[5]). In both cases, the (electronic) structure of the BL is a determinant of the graphene properties: in EM, the rippling of the BL determines the corrugation of the graphene sheet[6] as observed in Scanning Tunneling Microscopy (STM)[7,8]. In QFSMG, the interaction with the BL is very weak, but vacancies in the intercalation layer produce localized states, appearing located in regular patterns[9,10] which were proposed to descend from the BL symmetry[11]. Finally, the n-doping observed in EM was attributed to the effects of the surface states of the BL[12].

The conformation and electronic structure of the BL is therefore of primary importance, and somehow debated in the literature. Atomically resolved microscopy analyses at high temperature[2] reveal a pattern of corrugation appearing organized in crests forming an honeycomb-like lattice of nano-sized edge, whose periodicity was recognized as a "quasi" 6×6 (of SiC). However, analyses at low temperature and different voltages[3,13,14] reveal different electronic structures, appearing more sharp and localized, and different possible symmetries, including the 6√3×6√3 R30, which is the smaller consensus supercell between graphene and SiC when the two lattices are rotated of exactly 30 deg. This supercell is considered the "standard one" in several theoretical studies on BL, EM or QFSMG[15,16,17,18,19]. In some of these, the BL structure was studied and actually revealed a system of crests of ~1Å high, in agreement with the experimental observations. For QFSMG,[10,11], we recently proposed a alternative symmetry, namely √31×√31 R8.95 (of SiC), obtained by a slight rotation of the two lattices, and leading to a 1/3 sized consensus supercell which allows extensive calculations.

At variance of those of EM and QFSMG, however, the electronic properties of the BL alone received little attention: only very recently an ARPES[20] measurement has revealed the presence of a small band gap (~0.5eV) whose origin was not clearly attributed. The reason of the scarce interest towards this system is that it is usually considered only a support system to generate the most interesting one, that is, graphene.

In this work we adopt a different point of view, focusing on the BL itself. We are moved by two aims. The first is to clarify details of the structure of this system, since this influences the structure of EM and QFSMLG and can shed light on the kinetics of their formation processes, still basically unknown. The second is the analysis of the electronic structure of the BL as a system *per se*, to envision whether the specific properties of the BL could be exploited for applications. Our investigation tool is Density Functional Theory (DFT). The calculation scheme is reviewed in the next section, also including a description of the model systems, especially the non-standard ones, used for here the first time with BL. The subsequent section illustrates the results. Summary and discussion on perspectives for applications are reported in the last section.

## 2. Models and methods

### 2.1 Model Systems

The two model systems analyzed in this work are reported in Fig 1. The first one is the 6√3×6√3 R30 of SiC (or 13×13 of graphene, Fig 1, top part), considered the "standard" simulation cell for the SiC/graphene systems (hereafter called "L"). We included four SiC layers of the cubic polytype, and saturated the last one with H. In this work we also considered a smaller ("S") supercell recently used and validated for the QFSMG[10,11], the √31×√31 R8.95 of SiC (or 7×7 R21.787 of graphene, Fig 1, bottom part). In this case, the consensus between graphene and SiC supercells is obtained by allowing a small relative rotation (< 1 deg with respect to that in L model) of the two lattices and slightly different contraction of the buffer: in S model the surface density of the buffer atoms is 0.9% larger than in the L model. In the S supercell the BL can have two different stacking conformations[11] with respect to SiC, allowing respectively one site of type "hollow", i.e. with a Si atom lying under the center of a BL hexagon (Sh model) and one site of type "top", i.e. with Si atom lying under a C atom (St model, see red circles in Fig 1). These models are built locating a flat hexagonal carbon layer on the top of a neatly cut SiC surface at a distance slightly larger than the presumed bond distance and then relaxing (see next section for calculation details). The S supercell includes less than 1/3 atoms of the L one, reducing the computational of roughly one order of magnitude.

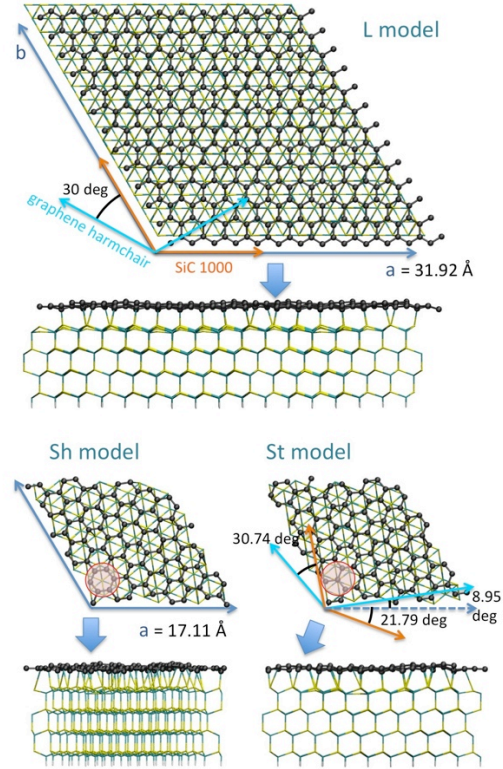

**Fig 1.** Model systems used in this work. Top and side views of the large (L) model and small models with "hollow" (Sh) and "top" (St) buffer stacking are reported (red circles indicate the hollow and top sites). The lattice vectors in the xy plane are represented as blue arrows, and their values reported. Orange arrows indicate the crystallographic directions of the SiC lattice, cyan arrows those of graphene, their relative rotation is reported. The thick arrows indicate the projection directions of the side views. L model includes 1310 atoms, and S models include 377 atoms each. Color coding: black = Carbon of the BL, cyan = carbon of SiC, yellow = Si, white = H. The views show the structurally optimized geometries.

## 2.2 Calculations Setup

DFT calculations are performed with a setup previously tested on this kind of systems[8]. The Rappe-Rabe-Kaxiras-Joannopoulos ultrasoft (RRKJUS) pseudopotentials[21] are used with the Perdew-Burke-Ernzerhof (PBE) exchange-correlation functional[22], van der Waals interactions are treated within the semi-empirical Grimme D2 scheme[23] (PBE-D2). The plane wave cutoff energy was set at 30 Ry and the density cutoff at 300 Ry, while the convergence threshold for self-consistency was set at $10^{-8}$. The BFGS quasi-Newton algorithm was used for structural optimizations[24], with standard convergence criteria, i.e. $10^{-3}$ a.u. for the forces and $10^{-4}$ a.u. for the energy. Γ point only is used for structural optimization, while for calculations of the density of states (DoS) we used a 10×10×1 grid generated using the Monkhorst and Pack scheme[25] for S supercells and a 5×5×1 grid for L. A Gaussian smearing of 0.01 Ry was set in all calculations. The lattice vectors of the supercells in the xy plane (reported in Fig 1) are chosen in such a way to have a relaxed SiC crystal. The third supercell parameter, in z direction, is the same in all cases (31.8Å), chosen large enough to decouple the periodic copies. The calculations were performed using Quantum ESPRESSO[26] (QE, version 5.3.0). Numerical data generated during this study are available from the corresponding author upon reasonable requests.

## 3. Results

In this section the results of this study are reported. The rippling pattern of the L model (part of section 3.1), were previously described by us[8] and others[6]. However to our knowledge a deep-in characterization of the structural features of the buffer in L or in the alternative models S was never reported. We believe that it is important to fill this gap, especially for the interpretation of energetic and electronic data, reported in the subsequent subsections.

### 3.1 General structural properties

The main structural properties are summarized in Fig 2. As it can be seen from panel (a), the structure of SiC lattice is rather regular in the bottom layer, and becomes less regular in the layer nearer to the surface, as indicated by the larger spread of the z coordinate distributions. The BL is located at an average distance of 2.4 Å from the Si layer, and displays a large vertical spread due to its rippling. The ripples are organized in specific patterns, clearly visible in panels (b)-(d) of Fig 2 (bright areas are protruding). In L the "crests" form a honeycomb lattice (profiled in yellow in panel (b)) separating three types of hexagonal irregular "tiles". This pattern is compatible with that observed by STM[2,20], although in those experiment the crests appear smeared. This could be attributed to the thermal fluctuations, not included in these calculations. In addition, it should be considered that STM images depend on voltage, whose variation can produce different sharpness and contrast[3,13,14]. Indeed, however, the experimental images the hexagonal "tiles" appears somehow more regular than those in L model. Interestingly, the corrugation pattern arising from model Sh, here first reported, is more similar in this respect, displaying only one kind of regular tile. The pattern of model St is instead formed by distorted hexagons, appearing "broken" on two sides. This situation appear similar to some what observed in some region of the STM of ref [2], where the crests seem to form rather parallel zig-zag lines weakly connected.

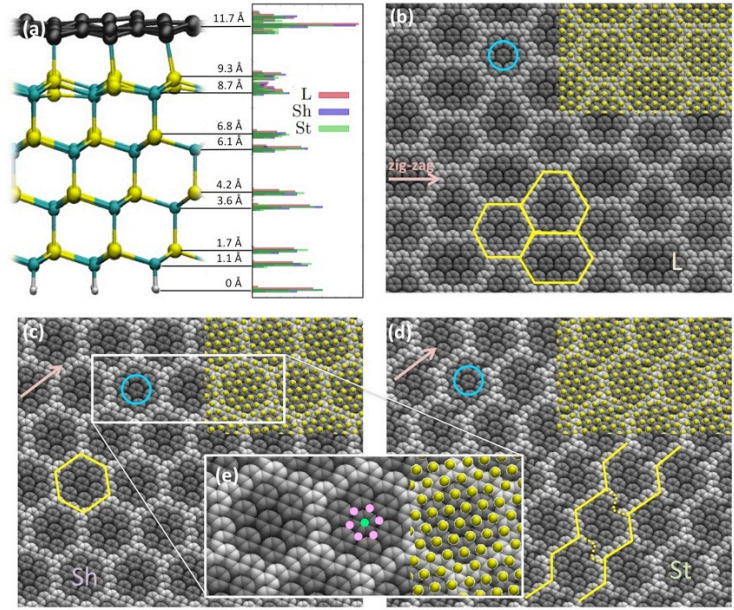

**Fig 2.** Structure of the buffer layer. (a) A side view of the system (specifically L model) with the average z coordinate of the layers reported. On the right, with the same scale, the distribution of the z coordinates of all atoms are reported, for the L, Sh and St model separately (color coding in the plot). Panels (b) to (d) report top views of the buffer layer (in vdW spheres representation), colored according to the z coordinate (bright grey protruding areas, dark=intruding). Yellow lines follow the profile of protruding areas. The cyan circles highlight the groups of benzene-like rings. Superimposed as yellow balls in the top right corner are the locations of the upper layer Si atoms. The inset (e) shows a zoom-in of the white profiled area in panel (c). In (e), location of Si atoms are also indicated as green and pink dots. Pink arrows in panels (b)-(d) indicate the zig-zag direction of graphene.

Locally, the BL appears organized in benzene-like units (highlighted by blue circles in panels (b-d)). In the zoomed inset (panel (e)) the crests appear formed by interconnected chains of those (broken) units, while the tiles include six, seven or eight benzene units, separated by intruding C atoms (colored in black and indicated by the pink dots in (e)). By comparing with the location of the Si atoms of the upper layer (yellow balls), it can be seen that each of the intruding C atom is located on top of a Si atom. As it will be shown in the next sections, they are actually covalently bound. However, not all the Si atoms are bonded to C. For instance, the central Si atom colored in green in panel (d) is found in a hollow position i.e. in the center of the benzene unit, and it is unbound. Top and hollow position roughly alternate, but the mismatch between the lattices unregisters the stacking. This modulated alternating of favorable (top) and unregistered relative C-Si positions creates the specific patterns of bonds of crests and benzene sub-units.

### 3.2 Bonding patterns

Fig 3 (a), reports the z profiles evaluated along different zig-zag lines in the three models shown in Fig 3 (b-d). The Si bonded C atoms are easily identified as those pointing downward, and are regularly separated by groups of C atoms belonging to the benzene units; the crest areas can be recognized as those pointing upwards. As it can be seen, the profiles of the L model (black and blue line in the bottom and top plot respectively) have the super-cell periodicity of ~3.2 nm, while the profiles of S models (magenta, green and red) are only approximately periodic, since the zig-zag line is not perfectly aligned to the crests, due to the rotation angle between substrate and BL. Nevertheless, the crests structures of the three models are locally superimposable (black vs magenta lines, belonging to L and Sh respectively, and red vs blue lines, belonging to L and St). The green profile is the most different of all, because it crosses the "broken crest", which is a peculiarity of the St model.

The covalent bonds between BL and the substrate can be identified in different ways: (i) by selecting the buffer layer atoms pointing downwards, or (ii) those at short distance from a Si atom, or (iii) by searching for the effective charge localization between buffer and substrate. Method (i) is the simplest but ambiguous, because the tail of distribution of z coordinates of bound and unbound atoms superimpose. Method (ii) identifies better the bound atoms, being the bond length distributed at ~2Å. Finally, method (iii) gives an immediate idea of the bonding pattern: Fig 3 (f-h) reports a representation of the total charge density evaluated on a plane located mid-way between the buffer and the substrate, as shown in (e). Bonds can be identified as charge density accumulation points (bright spots), located below a sub-set of the BL atoms. These are colored in pink in the superimposed image, and as anticipated, are distributed on a specific hexagonal pattern, alternating with the "benzene" rings (in cyan) appearing as grey hexagon in (b)-(d) and (f)-(h) due to their π orbitals structure. (The atoms of the crests appear dark in the image because they are unbound and far above the representation plane.) Fig 3 panels (f-h) also confirm the local resemblance of patterns of model L to that of model Sh and St. A comparative analysis of the three methods is reported in the SI.

### 3.3 Formation energies

The evaluation of relative energies of L and S models is not straightforward, because they have different sizes. One possibility is to normalize to the area of the cell, or equivalently, to the number of surface Si atoms. However, the surface density of atoms of the carbon layer is higher in the S model of ~1% (and, correspondingly, the corrugation level is of 2% larger see Table 1, two last columns). Therefore, a second non-equivalent normalization is to the number of atoms the BL. Even the evaluation of the total energy itself is non trivial. The global stability could be evaluated considering the formation energy $E_f$, i.e. the energy with respect to the sum of energy of all isolated atoms

$$E_f = E_{opt} - (N_H E_H + N_C E_C + N_{Si} E_{Si})$$

When renormalize to the unit surface, $E_f$ can be used as a measure of the relative stability, although it is not very meaningful *per se*, being dependent on the number of layers of bulk included in the model. The binding/dissociation energy of the BL was also evaluated, but it is less appropriate considering that the synthesis proceeds from the bulk (details in the SI).

The analysis of the energies indicates that Sh model is more stable than St model of about 0.01eV per surface Si atom. This is likely to be due to a more symmetric conformation of the bonding pattern, since the concentration of C atoms and the level of corrugation is basically the same in St and Sh. Additionally, S models turns out more stable of L of ~4.8 eV/Si surface atom. Because the main difference between S and L is the surface C density of the buffer, this points to a preference of the system for a slightly larger concentration, such as in S model. In general, however, the stabilization is likely to arise from the interplay between the different corrugation level (see Table 1) and the different arrangement/symmetry of the Si-C bonds and of the crests. Conversely, when one considers the formation energy per C buffer atom, the trend is inverted and L results lower than S of about 3.9 eV/ C atom. On the formal level, this is trivially due to the different surface concentration of C atoms in the two models and seems to indicate that the BL is more relaxed within the L model.

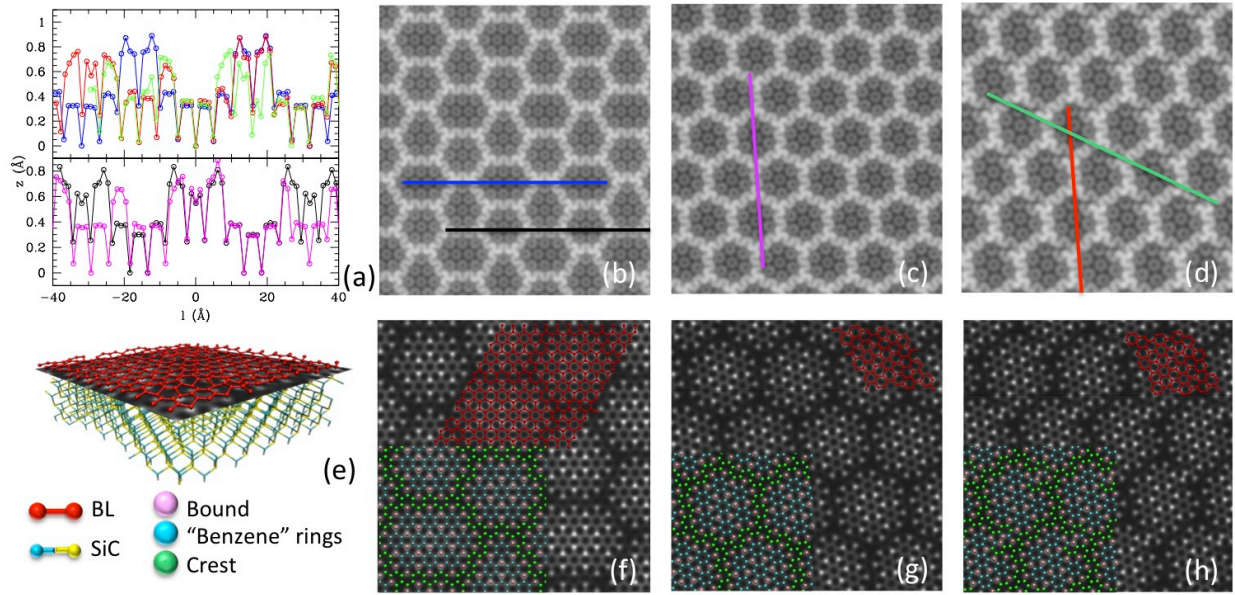

**Fig 3.** (a) z profiles of the BL atoms along the zig-zag lines depicted in panels (b)-(d), for the models L (b), Sh (c) and St (d) (colors of lines and plots correspond). The profiles are built including all the atoms of a single zig-zag line, the l coordinate being the projection of x-y coordinate onto this line. This imply that the separation of dots along the l coordinate of ~1.23Å. The AFM-like images (b)-(d) are obtained from the iso-density surfaces of the total electronic charge density, colored according to the eight (bright protruding, dynamical range of coloring = 1Å). In panels (e)-(h) the charge density is evaluated on a plane located between the buffer layer and the substrate, as shown in panel (e). Panels (e-f) report the value of the charge density on the plane (bright high density, dark low density), for models L (f), Sh (g) and St (h). Superimposed in red is the BL structure, and in colors the atoms of BL of the three different types: bound to SiC, "benzene-like" rings and crests, colored as shown in the color coding legenda in panel (e).

|    | $E_f$ (eV) | $N_{Si}$ | $N_C$ | $N_b$ | $N_b/N_C$ | $N_C/S$ (nm$^{-2}$) | $<\varphi^2>^{1/2}$ (deg) | $E_f/N_{Si}$ (eV) | $E_f/N_C$ (eV) |
|----|-----------|----------|-------|-------|-----------|---------------------|---------------------------|-------------------|----------------|
| L  | -180738.25 | 108 | 338 | 83 | 24.6% | 38.306 | 14.8 | -1673.5024 | -534.7286 |
| Sh | -52026.293 | 31  | 98  | 24 | 24.5% | 38.654 | 15.1 | -1678.2675 | -530.8805 |
| St | -52026.000 | 31  | 98  | 25 | 25.5% | 38.654 | 15.2 | -1678.2581 | -530.8776 |

**Table 1** Formation $E_f$ and binding energies $E_b$ of the three model systems as defined in the main text, and normalized to the to the number of Si surface atoms, of C buffer atoms and to the number of Si-C bonds, also reported. The average surface density of the buffer atoms, $N_C/S$ and the average corrugation level of the buffer, as measured by the standard deviation from zero of the out of plane dihedral are also reported $<\varphi^2>^{1/2}$.

### 3.4 Electronic properties

The electronic Density of States (DoS) for all models are reported in Fig 4. Once aligned to the Fermi level, they show the same global structure, displaying a low DoS area ~0.7 eV large, delimited by two rapidly increasing wings. This is an electronic gap of size comparable of that reported in ref [20], although a n-doping is present being the Fermi level located above the gap. A zoom in of the DoS (lower plot Fig 4) reveals some differences between the three models. Some residual DoS is present in the gap, revealing some more or less pronounced or separated "peaks" in the three models. The local DoS integrated in the gap range is reported as insets in Fig 4 (iso-density surfaces, in red), and clearly reveals that in-gap states are localized over the crests in the three cases. At a first sight, these states appear to be the π orbitals of the protruding C atoms. For the St case, where two separate peaks are clearly visible, we integrated over them separately (intervals indicated in yellow and purple in the plot). The inset on the right shows in corresponding colors the separated local DoS, and reveals that the lower energy in-gap peak (yellow) belong to states located on the more irregular sites, while the states belonging to higher energy in-gap peak (magenta) follow the more regular hexagonal pattern, which explains why in the Sh case only one peak (the "regular" one) is present, being the symmetry higher.

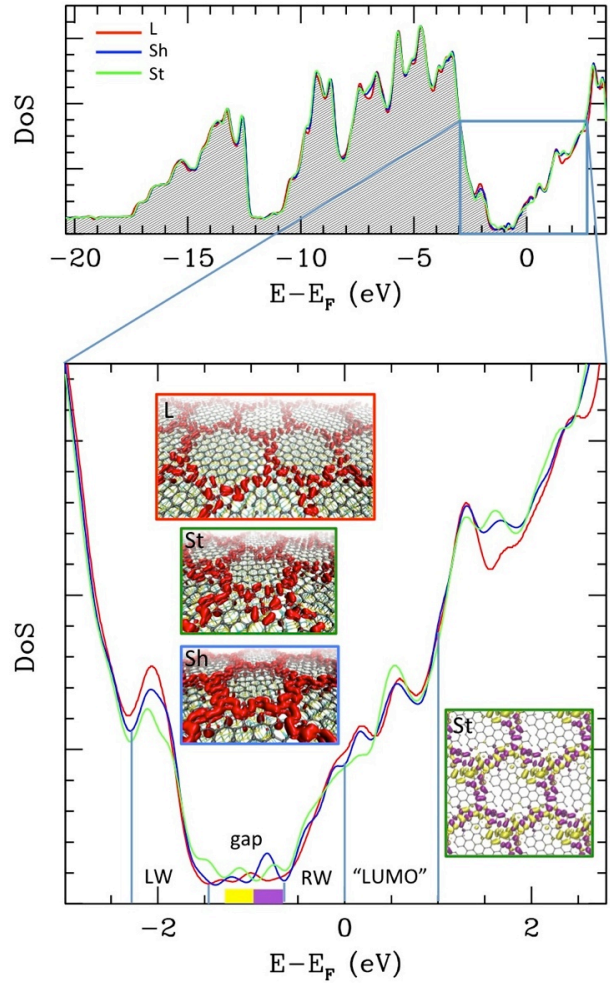

**Fig 4.** Total electronic DoS the L and S models. In the upper plot, the total DoS in the full energy range is reported for the three models (color coding reported) aligned to the Fermi Level. The shaded area includes the filled states. In the lower plot a zoom around the Fermi level is reported (same color coding for the lines). The three central insets report in red an iso-surface representation of the Local DoS integrated within the "gap" range [-0.7;-1.4] (the models are indicated and the insets are profiled with the same color coding of the lines); in the side inset profiled in green, the DoS of the St model integrated over the two separated in-gap peaks in the intervals indicated in yellow and purple in the plot are reported separately (in corresponding colors).

We systematically performed this analysis in other four energy intervals, namely the left wing (LW) and the right wing (RW) of the gap up to the Fermi level, and the the "LUMO" state, including a the lower unoccupied states up to 1eV (boundaries of the intervals indicated by vertical lines in Fig 4, lower plot). The local DoS of these intervals (and of the gap) are reported in Fig 5 in different representations aimed at showing the vertical localization of the states: The first two representations ("STM" and "vol slice over") display the electronic density on top of the buffer layer, and could be compared with STM images in fixed current and fixed height mode, respectively. The third one ("vol slice under") is obtained plotting the density in a plane located between the buffer and the substrate, and displays the inner part of the charge density. This analysis confirms that the in-gap states (second column from the left) are basically localized in the crests area: the superficial images show density localized on the protruding C atoms of the crests, while the inner image shows density localized under the C atoms intruding, i.e. those covalently bound to the substrate. Overall, then, the in-gap states represent bonding orbitals either between the crests atoms or between the intruding crests atoms and the substrate. The in-gap states do not display charge density on the "benzene" units.

Conversely, the distribution of the LW states is localized on the benzene units, or under them, as shown by the "slice under" images. Besides those clearly visible rings, the inner images show also density localized under the C sites interstitial between rings, which are bonded to the substrate. Also the superficial image shows density on the benzene rings, visible as small circles within the larger hexagonal shapes delimited by crests, besides some residual density of the $\pi$ orbitals over the protruding atoms. Overall, these states appear to have bonding character, either between the out of crest atoms, or between them and the substrate.

The charge density of the states of the RW below the Fermi level has a strong inner component under the "tiles", which appear however to be different, being localized over Si atoms located in the hollow positions, and therefore these states do not participate to the covalent bonding to the substrate, but are rather "dangling Si bonds". The superficial charge density is localized mainly on the crests, but appears to be a sum of localized $p_z$ orbitals rather than a $\pi$ system. These two aspects are similar and enhanced in the "LUMO" states: the superficial states are protruding $p_z$ orbitals and the inner states are dangling bonds under the "benzene rings" (although with a different symmetry with respect to the RW states). Overall, then, the states above the gap appear to have an increasing anti-bonding character.

4. **Discussion and conclusions**

In this work we have comparatively analyzed results from three different model systems for the buffer carbon layer on the Si-rich surface of SiC: the "standard one" (L) with exactly 30 deg rotation between the SiC and buffer lattices, 13×13 / 6√3×6√3 supercells and a superficial C buffer atoms density of ~38.3 C atom/nm$^2$, and two alternative ones (Sh and St with 7×7 / √31×√31 supercells), with a slightly different rotation between the two lattices (~29 deg), slightly larger superficial C density (~38.7 C/nm$^2$). L is fully compatible with the (quasi) 6×6/SiC supercell, while S models are compatible only approximately.

The three models share common general structural features: a system of crests of alternating protruding and intruding C atoms forming an honeycomb pattern and separating hexagonal "tiles" of ~1nm side. Within the "tiles" carbon organizes in "benzene-like" rings, separated by intruding C sites. These and the intruding sites in the crests are covalently bound to Si atoms of the substrate. The main differences between the three models are in the form of the "tiles": these are regular hexagons in the Sh model, distorted hexagons with partially broken sides in St, while in the L model three different types of hexagons, one regular and two distorted are present. The comparison with the atomic microscopy data does not completely solve the ambiguity between the models: AFM and STM images clearly show the honeycomb system of crests of size compatible with all of them. Small irregularities of the tiles seem to appear in some images[20,13], while in others they are not visible[2]. In general, however, crests appear smeared and regularized with respect to L model. While both the effects could be attributed to the already mentioned thermal fluctuations (neglected in our calculations), the regularization is an indication in favor of Sh model.

Some features of the fine structure we observe, e.g. the "benzene-like rings", are barely reported in the literature. However, we observe that these need very high resolution and specific voltage conditions to be revealed. In fact, structures that resemble those of Fig 5, with modulation of contrast on the crests and within the tiles similar to benzene rings were observed in ref [14], where a systematic study at different voltages and temperature was performed. In summary, the theory-experiment comparison of structural data would indicate that none of the models should be excluded.

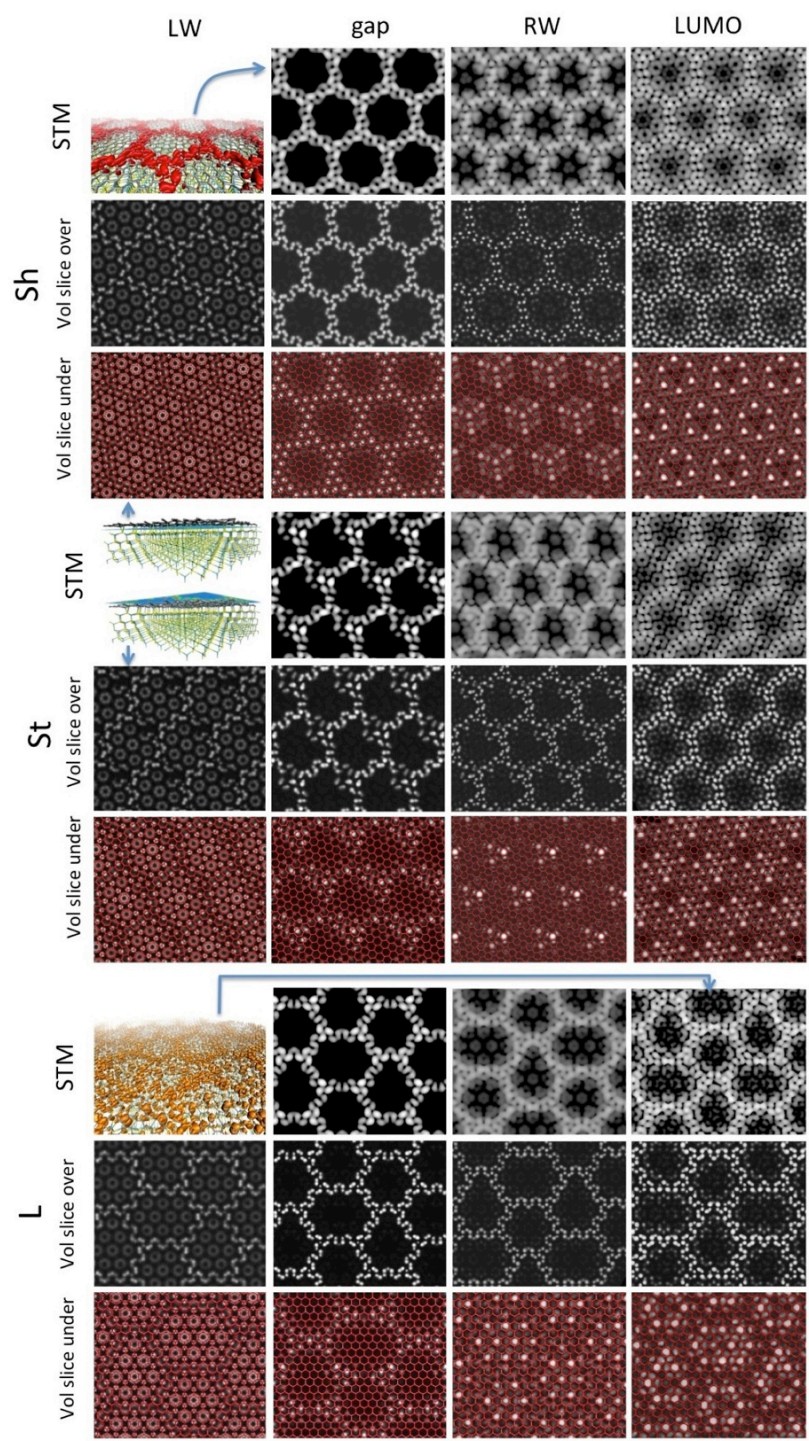

**Fig 5.** Local DoSs integrated over different energy intervals (columns: LW= left wing [-2.3;-1.4]; gap [-1.4; -0.7]; RW= right wing [-0.7;0]; "LUMO" [0;+1]) for the three different models (from top to bottom: Sh, St and L, each represented in three different modes: STM = "fixed current" mode, obtained coloring an iso-density surface (~$10^{-4}$ a.u.) according to the height; "vol slice over" obtained coloring a plane placed over the layer according to the local value of the density, and "vol slice under", same but with the plane placed between the buffer – reported in red in wireframe representation – and the substrate (the location of the two planes is represented in the forth row, first cell). Also represented are an iso-density surface of the "gap" local DoS (in red, first row, first cell) and of the LUMO state (in red, $7^{th}$ row, first cell).

Other indications come from the analysis of stability. The comparison between L and S models indicates that the formation energy per unit surface is smaller for the S models (the smallest for Sh), while the same evaluated per C atom of the BL is smaller for L. This is not a contradiction, however, because the superficial concentration of C atoms of the buffer is larger in S models. It might seem not intuitive that a more crowded (and consequently more corrugated) conformation is more stable than a more "diluted" (and slightly flatter). However, one must consider that the buffer is not a completely flat $sp^2$ graphene, nor even a completely $sp^3$ hybridized graphane-like sheet, but a combination of the two, whose optimal corrugation/contraction level is not easy to evaluate *a priori*. Therefore it is not straightforward to estimate which is the optimal superficial density of C atoms in the buffer. In addition, Sh conformation might be structurally favored by its high symmetry, bringing a particularly favorable binding pattern. This conclusion is also compatible with a recent general theoretical analysis showing that conformations with relative rotations of lattices with low commensuration might help releasing the strain[27].

Our results for the electronic structure return similar DoS for the three models, displaying a gap approximately 0.7 eV wide, with localized in-gap states and n-doping. We remark that, while a similar n-doping level is reported in the theoretical calculations (e.g. [20]), this is not fully confirmed by experiments, which seem rather to indicate the Fermi level located in proximity of the gap or within it. This may be attributed to different causes: extrinsic doping induction from the experimental setup or poor treatment of the bulk polarization in the theoretical calculations [11,20]. However, the doping level is not likely to affect the features (shape, localization, extension) of the electronic states, which in fact, appear to be fully compatible with the structural analysis. The basically non-dispersive in-gap states appear to descend from the π orbital system of the crests atoms and are fully localized on them, constituting the binding system between crests atoms or among them and the substrate. The specific space distribution of those state of course depends on the model (being the crests different in the three models) and in particular the symmetry breaking occurring in Sh seems to split the in-gap state in two distinct ones, with separate spatial distribution, while this splitting is not observed in the more symmetric Sh model.

Furthermore, just below the gap, we observe a more conventional π binding system, localized in the "benzene like" rings within the "tiles", though with some residual density on the crests. On the other side, above the gap, we revealed a set of dangling Si bonds localized underneath the buffer layer, and a set of $p_z$ orbitals mainly localized over the crests and in proximity of them. These states might have relevance in the intrinsic doping of the graphene monolayer on BL, when present, which was attributed to donor surface states either of the buffer or of the SiC surface[12]. The experimental observation of this variety of states requires a systematic study STM analysis of the BL at different voltages, performed only in a few studies, which in fact, reveal the presence of structures we observe in our analysis, such as the already mentioned benzene rings[14], localized pz orbitals or dangling bonds[13], often organized in triangular or hexagonal symmetries, compatible with our models. In those images both regular and irregular hexagonal "tiles" are observed, as far as areas with broken crests typical of St model.

In summary, our work does not completely solve the problem of which is the exact symmetry and conformation of the BL, conversely suggesting that different symmetries might coexist and possibly inter-convert at sufficiently high temperature. The symmetry of L gives a more extended spatial commensurability, a relative rotation that seems globally more compatible apparently exact 30 deg rotation of the two lattices. However S models seems to be locally more stable. The preference for one or the other conformation should be investigated accounting also for the kinetics of the buffer formation, which is currently under investigation and matter of a forthcoming paper[28]. From the modeling perspective, S models are much less expensive, and certainly a good approximation of the real sample.

The peculiar structural and electronic properties of the BL suggest interesting applications. The corrugation and the considerable variability of electronic structure between crests and "tiles" suggest that this system is particularly reactive[29]: states localized in the different energy regions (gap, and its sides) display well separate spatial localization and very different (sometimes opposite) "chemical" character. The correlation between these two features suggests that different chemical species (e.g. electrophilic, nucleophile, or dienophile) might select different spatial areas, following the moiré superlattice, offering a unique potentiality of obtaining a nano-patterned chemical functionalization.

**Acknowledgments**

We thank Dr. Stefan Heun, Prof. Paolo Giannozzi, Dr Luca Bellucci, Dr. Camilla Coletti, Dr. Yuya Murata, and Dr. Vittorio Pellegrini for useful discussions. We gratefully acknowledge financial support by EU-H2020, Graphene-Core1 (agreement No 696656) and Core2 (agreement No 725219), by CINECA awards IsB11_flexogra (2015), IsC36_ElMaGRe (2015), IsC44_QFSGvac (2016), IsC44_ReIMCGr (2016) and PRACE "Tier0" award Pra13_2016143310 (2016). We acknowledge CINECA staff for technical support.

**Author contributions**

TC performed calculations, produced data and performed part of the analyses. VT finalized the analyses, produced the figures and wrote the manuscript.

**Additional information**

Competing interest: the authors declare no competing interests.

# References

[1] Norimatsu, W.; Kusunoki, M. Epitaxial Graphene on SiC{0001}: Advances and Perspectives. *Phys. Chem. Chem. Phys*. **16,** 3501 (2014)

[2] Goler, S.; Coletti, C.; Piazza, V.; Pingue, P.; Colangelo, F.; Pellegrini, V.; Emtsev, K. V.; Forti, S.; Starke, S.; et al. Revealing the Atomic Structure of the Buffer Layer between SiC (0001) and Epitaxial Graphene. *Carbon* **51**, 249−254 (2013)

[3] Riedl, C.; Coletti, C.; Starke, U. Structural and Electronic Properties of Epitaxial Graphene on SiC(0001): a Review of Growth, Characterization, Transfer Doping and Hydrogen Intercalation. *J. Phys. D: Appl. Phys*. **43**, 374009 (2010)

[4] Fiori S, Murata Y, Veronesi S, Rossi A, Coletti C, and Heun S Li-intercalated graphene on SiC(0001): An STM study *Phys. Rev. B* **96**, 125429 (2017)

[5] Riedl C., Coletti C., Iwasaki T., Zakharov A. A., and Starke U. Quasi-Free-Standing Epitaxial Graphene on SiC Obtained by Hydrogen Intercalation *Phys. Rev. Lett.* **103**, 246804 (2009)

[6] Mallet, P.; Varchon, F.; Naud, C.; Magaud, L.; Berger, C.; Veuillen, J. Y. Electron States of Mono- and Bilayer Graphene on Si Probed by Scanning-Tunneling Microscopy. *Phys. Rev. B*:. **76**, 041403 (2007)

[7] Telychko, M.; Berger, J.; Majzik, Z.; Jelínek, P.; Švec, M. Graphene on SiC(0001) Inspected by Synamic Atomic Force Microscopy at Room Temperature. Beilstein *J. Nanotechnol*. **6**, 901−906 (2015)

[8] Cavallucci T and Tozzini V Multistable Rippling of Graphene on SiC: A Density Functional Theory Study *J. Phys. Chem. C*, **120**, 7670−7677 (2016)

[9] Murata Y, Mashoff T, Takamura M, Tanabe S, Hibino H, Beltram F, and Heun S, Correlation between morphology and transport properties of quasi-free-standing monolayer graphene *App Phys Lett* **105**, 221604 (2014)

[10] Murata Y, Cavallucci T, Tozzini V, Pavliček N, Gross L, Meyer G, Takamura M, Beltram F, Hibino H, Heun S. Atomic and Electronic Structure of Si Dangling Bonds in Quasi-Free-Standing Monolayer Graphene *Nano Res* **11** , 864–873 (2018)

[11] Cavallucci T, Murata Y, Takamura M, Hibino H, Heun S, Tozzini V, Unraveling localized states in quasi free standing monolayer graphene by means of Density Functional Theory *Carbon* **130**, 466-474 (2018)

[12] Ristein J., Mammadov S., Seyller Th. Origin of Doping in Quasi-Free-Standing Graphene on Silicon Carbide *Phys Rev Lett,* **108,** 246104 (2012)

[13] Riedl C "Epitaxial Graphene on Silicon Carbide Surfaces: Growth, Characterization, Doping and Hydrogen intercalation" PhD Thesis (2010), Fridrich Alexander Universität Erlangen, Nurberg

[14] Hu T. W., Ma F., Ma D. Y., Yang D., Liu X. T., Xu K. W., Chu P.K. Evidence of atomically resolved 6x6 buffer layer with long-range order and short-range disorder during formation of graphene on 6H-SiC by thermal decomposition App Phys Lett, **102,** 171910 (2013)

[15] Kim, S.; Ihm, J.; Choi, H. J.; Son, Y.-W. Origin of Anomalous Electronic Structures of Epitaxial Graphene on Silicon Carbide. Phys. Rev. Lett. 100, 176802, (2008)

[16] Varchon, F.; Mallet, P.; Veuillen, J.-Y.; Magaud, L. Ripples in Epitaxial Graphene on the Si-terminated SiC(0001) Surface. Phys. Rev. B: Condens. Matter Mater. Phys. 77, 235412 (2008)

[17] Sforzini J., Nemec L., Denig T., Stadtmüller B., Lee T.-L., Kumpf C., Soubatch S., Starke U., Rinke P., Blum V., Bocquet F. C., and Tautz F. S. Approaching Truly Freestanding Graphene: The Structure of Hydrogen-Intercalated Graphene on 6H-SiC(0001) Phys Rev Lett 114, 106804 (2015)

[18] Deretzis I. La Magna A. Role of covalent and metallic intercalation on the electronic properties of epitaxial graphene on SiC(0001) *Phys Rev B* **84**, 235426 (2011)

[19] Cavallucci T, Tozzini V Multistable Rippling of Graphene on SiC: A Density Functional Theory Study J of Phys Chemi C **120** 7670-7677 (2015)

[20] Nair M N., Palacio I, Celis A, Zobelli A, Gloter A, Kubsky S, Turmaud J-P, Conrad M, Berger C, de Heer W, Conrad E H, Taleb-Ibrahimi A, Tejeda A Band Gap Opening Induced by the Structural Periodicity in Epitaxial Graphene Buffer Layer *Nano*

*Lett.*, **17**, 2681−2689 (2017)

[21] Rappe, A. M.; Rabe, K. M.; Kaxiras, E.; Joannopoulos, J. D. Optimized Pseudopotentials. *Phys. Rev. B:* **41**, 1227 (1990)

[22] Perdew, J. P.; Burke, K.; Ernzerhof, M. Generalized Gradient Approximation Made Simple. *Phys. Rev. Lett.* **77**, 3865−3868. (1996)

[23] Grimme, S. Semiempirical GGA-type Density Functional Constructed with a Long-Range Dispersion Correction. *J. Comput. Chem.* **27**, 1787 (2006)

[24] Billeter, S. R.; Curioni, A.; Andreoni, W. Efficient Linear Scaling Geometry Optimization and Transition-State Search for Direct Wavefunction Optimization Schemes in Density Functional Theory Using a Plane-Wave Basis. *Comput. Mater. Sci*. **27,** 437. (2003)

[25] Monkhorst, H. J.; Pack, J. D. Special Point for Brillouin Zone Integration. *Phys. Rev. B* **13,** 5188. (1976)

[26] Giannozzi, P. QUANTUM ESPRESSO: a Modular and Open- Source Software Project for Quantum Simulations of Materials. *J. Phys.: Cond. Matt* **21**, 395502 (2009)

[27] Sclauzero G., Pasquarello A Low-strain interface models for epitaxial graphene on SiC(0001) *Diam & Rel Mat* **23** 178–183. (2012)

[28] Bellucci L, Cavallucci T, Tozzini V From buffer layer to graphene on Silicon Carbide: a multi-scale simulation, in preparation

[29] Cavallucci T, Kakhiani K, Farchioni R Tozzini V Morphing graphene-based systems for applications: perspectives from simulations, GraphITA. Carbon Nanostructures., 87-111 (2017)

# Intrinsic structural and electronic properties of the Buffer Layer on Silicon Carbide unraveled by Density Functional Theory

Tommaso Cavallucci and Valentina Tozzini

*NEST- Scuola Normale Superiore and Istituto Nanoscienze, Cnr, Piazza San Silvestro 12, 56127 Pisa, Italy*

# Supporting information

## S.1. Distributions of bonds

As mentioned in the main text, the covalent bonds distribution can be evaluated in different ways (i) by evaluation of the z coordinate (ii) by selecting the Si-C minimum distance for each C atom in the BL and (iii) evaluating the charge density value in a point midway between each C atom in the BL and the nearest Si atom (see Fig S.1 b). Fig S.1 (a) compares the methods (ii) and (iii) showing the distribution of Si-C minimal distances and the values of charge density in the midway point. The two sets of distribution (three for each method, for the three models) both display three separate populations, corresponding to the bound atoms (small Si-C distance, high density), the unbound atoms (intermediate distance and density) and crests atoms (large distance, small density). The populations selected with the two methods coincide and are represented in colors in Fig S.2, (a-c).

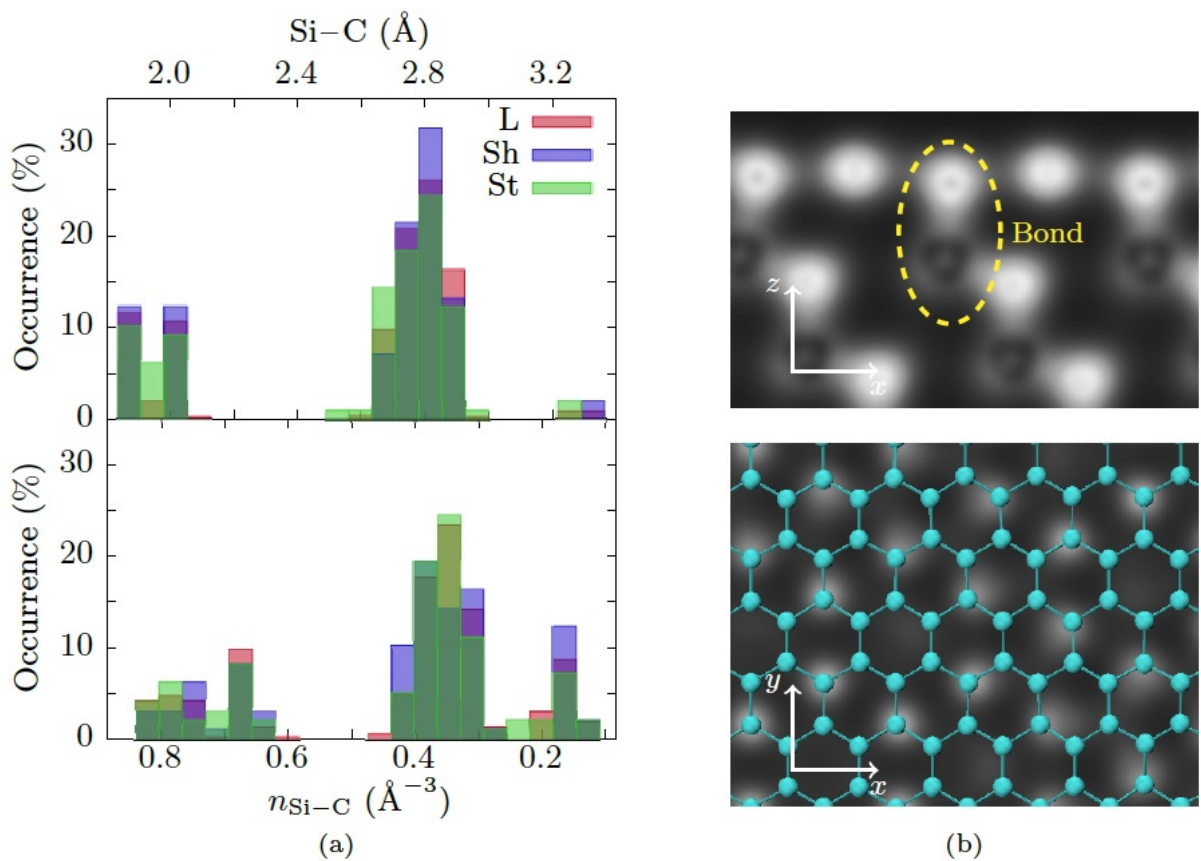

**Fig S.1** (a) Distributions of the minimum distances between C atoms of the BL and Si atoms of the substrate (top) and of the charge density value evaluated in the middle points of the Si-C distance from a C of the buffer and the nearest Si (the latter shown with inverted abscissa, to better compare with the fomer). (b) Total electronic charge density evaluated on a vertical plane cutting some Si-C bonds (indicated) and on an horizontal plane located between the BL and the substrate.

The comparison of methods (ii)-(iii) and (i) is reported in Fig S.2, (d)-(e). It can be seen that the population of bound and unbound atoms are not very well separable on the basis of the z coordinate, because their tail superimpose. The bonds spatial distribution is visible also in Fig S.3 as they appear as bright spots when the charge density is plotted on a plane between the BL and the substrate. In Fig S.3, also visible (less bright) is the aromatic π system of the "benzene" rings (corresponding to "unbound" atoms in dark grey in Fig S.2). As explained in the text, the rings are separated by bound atoms.

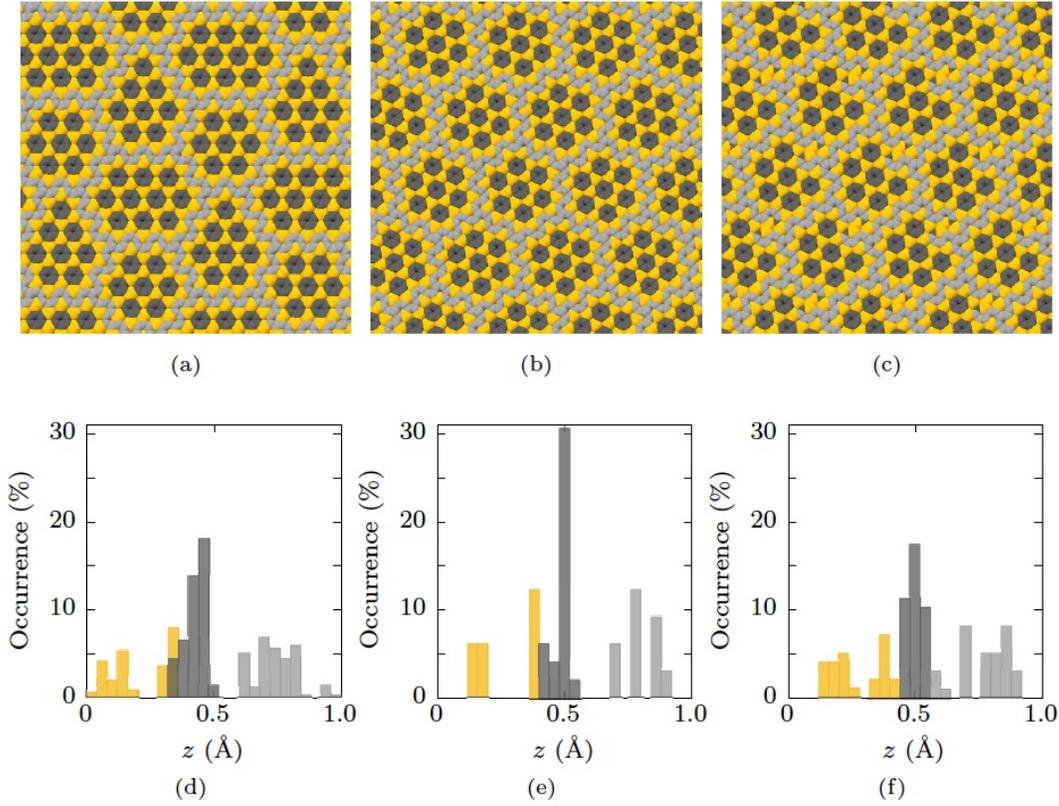

**Fig S.2** Space distribution of bound (yellow), unbound (dark grey) and crests atoms (light grey), in L (a), Sh (b) and St (c) models, as selected with the method (ii). (d-f) report the height distribution of the three classes of atoms, colored as in (a-c).

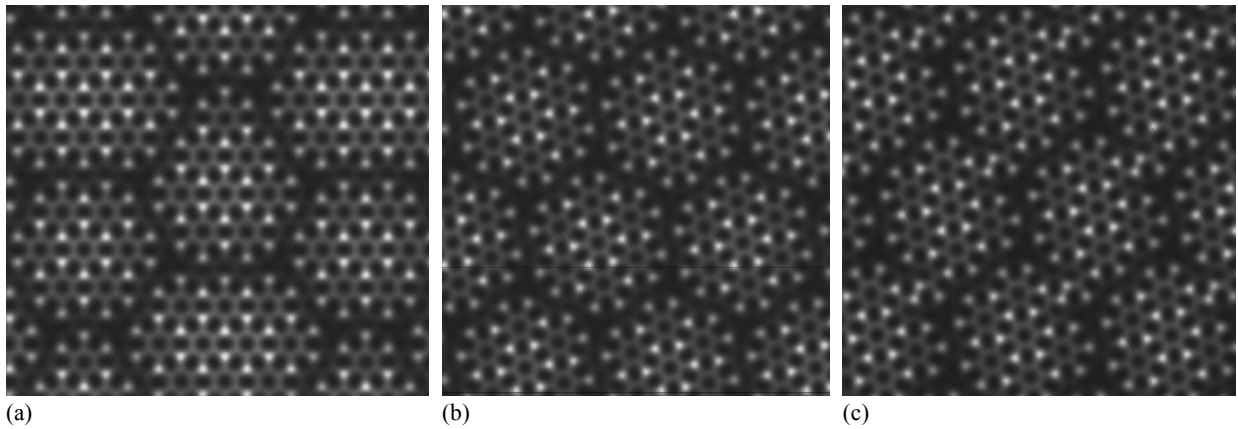

(a)  (b)  (c)
**Fig S.3** Charge density distribution plotted on a plane located between the BL and the substrate, as in Fig S.1 (b) (upper), for the three models (a)=L, (b)=Sh, (c)=St.

## S.2. Binding energy

An alternative to the formation energy Ef to evaluate the relative stability of the models is the binding energy, e.g. the energy of the system evaluated with respect to the graphene sheet separated from the substrate

$$E_b = E_{opt} - (E_{sub} + E_{gr})$$

being $E_{opt}$, $E_{sub}$ and $E_{gr}$ the energy of the systems and of their substrate and graphene isolated components respectively evaluated in the same cell. $E_b$ accounts both for the chemical energy included in the covalent bonds between the buffer and the substrate and of their structural readjustment due to binding. This quantity (and its value per unit surface, per Si atom, per C atom and per bond is reported in Table S.1.

For S models the estimated dissociation energy of the buffer to the separated graphene is ~0.7eV (little larger for Sh) per Si-C covalent bond, which is little less than what expected for an average value for a covalent bond. Conversely the value for L model is considerably larger, indicating that the sheet is more difficult to detach in the L model than in S. We remark, however, that this is just a very rough evaluation of the binding energy, especially because the reference system is not the flat graphene, but a laterally compressed graphene sheet. Therefore the reference systems in the two cases are different being the lateral compression different.

|    | $E_b$ (eV) | $N_{Si}$ | $N_C$ | $N_b$ | $N_b/N_C$ | $N_C/S$ (nm$^{-2}$) | $E_b/N_{Si}$ (eV) | $E_b/N_C$ (eV) | $E_b/N_b$ (eV) |
|----|------------|----------|-------|-------|-----------|---------------------|-------------------|-----------------|-----------------|
| L  | -199.667   | 108      | 338   | 83    | 24.6%     | 38.306              | -1.849            | -0.5907         | -2.406          |
| Sh | -16.870    | 31       | 98    | 24    | 24.5%     | 38.654              | -0.544            | -0.1721         | -0.703          |
| St | -16.577    | 31       | 98    | 25    | 25.5%     | 38.654              | -0.535            | -0.1692         | -0.663          |

**Table S.1** Binding energies $E_b$ of the three model systems as defined in the main text, and normalized to the to the number of Si surface atoms, of C buffer atoms and to the number of Si-C bonds, also reported. The average surface density of the buffer atoms, $N_C/S$ is also reported.